\input harvmac
\input epsf

\newcount\figno
\figno=0
\def\fig#1#2#3{
\par\begingroup\parindent=0pt\leftskip=1cm\rightskip=1cm\parindent=0pt
\baselineskip=12pt
\global\advance\figno by 1
\midinsert
\epsfxsize=#3
\centerline{\epsfbox{#2}}
\vskip 14pt

{\bf Fig. \the\figno:} #1\par
\endinsert\endgroup\par
}
\def\figlabel#1{\xdef#1{\the\figno}}
\def\encadremath#1{\vbox{\hrule\hbox{\vrule\kern8pt\vbox{\kern8pt
\hbox{$\displaystyle #1$}\kern8pt}
\kern8pt\vrule}\hrule}}

\overfullrule=0pt

\noblackbox
\parskip=1.5mm

\def\Title#1#2{\rightline{#1}\ifx\answ\bigans\nopagenumbers\pageno0
\else\pageno1\vskip.5in\fi \centerline{\titlefont #2}\vskip .3in}

\font\caps=cmcsc10

\noblackbox
\parskip=1.5mm

  
\def\npb#1#2#3{{\it Nucl. Phys.} {\bf B#1} (#2) #3 }
\def\plb#1#2#3{{\it Phys. Lett.} {\bf B#1} (#2) #3 }
\def\prd#1#2#3{{\it Phys. Rev. } {\bf D#1} (#2) #3 }
\def\prl#1#2#3{{\it Phys. Rev. Lett.} {\bf #1} (#2) #3 }

\def\cmp#1#2#3{{\it Commun. Math. Phys.} {\bf #1} (#2) #3 }

\def\bb#1{{\tt hep-th/#1}}

\def\jhep#1#2#3{{\it J. High Energy Phys.} {\bf #1} (#2) #3 }


           \def\CO{{\cal O}} 
   \def\CG{{\cal G}}
\def\CL{{\cal L}} \def\CH{{\cal H}}

\def\CN{{\cal N}}


\def\dj{\hbox{d\kern-0.347em \vrule width 0.3em height 1.252ex depth
-1.21ex \kern 0.051em}}

\def\half{{1\over 2}\,}
\def\d{{\rm d}}

\def\Tr{{\rm Tr\,}}

\def\ket{\rangle}
\def\bra{\langle}

\def\pt{\partial}

\def\bL{\overline L} 
\def\Dirac{\,\raise.15ex\hbox{/}\mkern-13.5mu D}
\def\dirac{\,\raise.15ex\hbox{/}\kern-.57em \partial}
\def\shalf{{\ifinner {\textstyle {1 \over 2}}\else {1 \over 2} \fi}} 
\def\sshalf{{\ifinner {\scriptstyle {1 \over 2}}\else {1 \over 2} \fi}} 
\def\sfourth{{\ifinner {\textstyle {1 \over 4}}\else {1 \over 4} \fi}}

\lref\rads{J. Maldacena, {\it Adv. Theor. Math. Phys.} {\bf 2} (1998)
231 \bb{9711200.}
 S.S. Gubser, I.R. Klebanov and A.M. Polyakov,
\plb{428}{1998}{105} \bb{9802109.} E. Witten, 
{\it Adv. Theor. Math. Phys.} {\bf 2} (1998)
253 \bb{9802150.} }

\lref\rsuspouno{L. Dyson, J. Lindesay and L. Susskind,  
\jhep{0208}{2002}{045}  
\bb{0202163.}}

\lref\rsuspodos{L. Dyson, M. Kleban and L. Susskind, 
\jhep{0210}{2002}{011}  
\bb{0208013.}} 

\lref\rsuspotres{N. Goheer, M. Kleban and L. Susskind, 
\jhep{0307}{2003}{056}  \bb{0212209.}} 

\lref\rsolo{D. Birmingham, I. Sachs,  S. N. Solodukhin,  
\prd{67}{2003}{104026}  
\bb{0212308.}} 

\lref\rbrick{G.  't Hooft,  
\npb{256}{1985}{727.}}       

\lref\rsusug{L. Susskind and J. Uglum,  
\prd{50}{1994}{2700}  
\bb{9401070.}} 
 
\lref\rberk{M. Berkooz, B. Craps,  D. Kutasov and  G. Rajesh, 
\jhep{0303}{2003}{031}  
\bb{0212215.}}
 
\lref\rlowe{N. Iizuka, D. Kabat, G. Lifschytz and  D. A. Lowe,  
\prd{67}{2003}{124001} 
\bb{0212246.}  \bb{0306209.}}  

\lref\rthirring{W. Thirring, {\it ``Quantum Mechanics of Large
Systems"} Springer-Verlarg, New York 1984.}

\lref\rsrednicki{M. Srednicki, {\tt cond-mat/9809360.}}

\lref\rsdollar{S.W. Hawking, \prd{14}{1976}{2460.}}

\lref\rpesk{T. Banks, L. Susskind and  M. E. Peskin,  
\npb{244}{1984}{125.} }

\lref\rHH{J. B. Hartle and S.W. Hawking, \prd{13}{1976}{2188.}}

\lref\rGH{G.W. Gibbons and S.W. Hawking, \prd{15}{1977}{2752.}}

\lref\rhordiv{J.L.F. Barb\'on,  
\prd{50}{1994}{2712}  
\bb{9402004.}}

\lref\rhaghor{J.L.F. Barb\'on,  
\plb{339}{1994}{41} 
\bb{9406209.}}
  
\lref\rroberto{J.L.F. Barb\'on and R. Emparan,  
\prd{52}{1995}{4527}  
\bb{9502155.}} 

\lref\ryork{J. W. York, Jr,   
\prd{33}{1986}{2092.}} 

\lref\rstro{J. Maldacena and A. Strominger,  
\jhep{9812}{1998}{005}  
\bb{9804085.}}

\lref\rshenk{P. Kraus, H. Ooguri and  S. Shenker,  
\prd{67}{2003}{124022}  
\bb{0212277.} L. Fidkowski, V. Hubeny, M. Kleban and S. Shenker,  
\bb{0306170.}}   

\lref\rmartinec{E. Martinec, \bb{9809021.}}  

\lref\rsfet{K. Sfetsos and K. Skenderis,  
\npb{517}{1998}{179} 
\bb{9711138.}}
 
\lref\renglert{F. Englert and E. Rabinovici,  
\plb{426}{1998}{269}  
\bb{9801048.} }
 
\lref\rcensor{J.L.F. Barb\'on and E. Rabinovici,  
\jhep{0203}{2002}{057}  
\bb{0112173.} 
{\it Found. Phys. } {\bf 33} (2003) 145 
\bb{0211212.}}

\lref\rstre{K.S. Thorne, R.H Price and D.A. McDonald, {\it ``Black Holes: The
Membrane Paradigm".} Yale University Press 1986.
 L. Susskind, L. Thorlacius and  J. Uglum,  
\prd{48}{1993}{3743} 
\bb{9306069.}}  
      
\lref\rbtz{M. Ba\~nados,  C. Teitelboim and   J. Zanelli, 
\prl{69}{1992}{1849}  
\bb{9204099.}}

\lref\rmathur{O. Lunin and  S. D. Mathur,  
\npb{623}{2002}{342}  
\bb{0109154.} \prd{88}{2002}{211303} \bb{0202072.}  
}  

\lref\rarnold{V.I. Arnold, A. Weinstein and  K. Vogtmann, {\it ``Mathematical Methods of 
Classical Mechanics".}, Springer Verlag (1989)}  

\lref\rmaldas{J. Maldacena, \jhep{0304}{2003}{021}  
\bb{0106112.}}

\lref\rsuscumple{L Susskind,  
Contribution to Stephen Hawking's 60th birthday celebration. 
\bb{0204027.}}   

\lref\rmarti{T. Banks, M. R. Douglas, G. T. Horowitz and E. Martinec,
\bb{9808016.}}

\lref\rusex{J.L.F. Barb\'on and E. Rabinovici, \npb{545}{1999}{371} 
\bb{9805143.}}

\lref\raps{A. Adams, J. Polchinski and E. Silverstein, \jhep{0110}{2001}{029}
  \bb{0108075.}}

\lref\rolddabho{A. Dabholkar, \npb{439}{1995}{650} 
\bb{9408098.} \plb{347}{1995}{222}  
\bb{9409158.}}   

\lref\rdabho{A. Dabholkar, \prl{88}{2002}{091301} \bb{0111004.}}

\lref\rthresholds{J.L.F. Barb\'on, I.I. Kogan and E. Rabinovici,
\npb{544}{1999}{104,} \bb{9809033.}}

\lref\rHPage{S.W. Hawking and D. Page, \cmp{87}{1983}{577.}}

\lref\rghen{G.W. Gibbons and S.W. Hawking, \prd{15}{1977}{2752.}}

\lref\rgibperry{G.W. Gibbons and M.J. Perry,  
{\it Proc. Roy. Soc. Lond.} {\bf A 358} (1978) 467.}
 
\lref\rgpy{D.J. Gross, M.J. Perry and L.G. Yaffe, \prd{25}{1982}{330.}}

\lref\rwithp{E. Witten, {\it Adv. Theor. Math. Phys.} {\bf 2} (1998)
505 \bb{9803131.}}

\lref\rmalda{J. Maldacena, {\it Adv. Theor. Math. Phys.} {\bf 2} (1998)
231 \bb{9711200.}}


\baselineskip=14pt

\line{\hfill CERN-TH/2003-167}
\line{\hfill RI-03/07-007}
\line{\hfill {\tt hep-th/0308063}}

\vskip 0.5cm

\Title{\vbox{\baselineskip 12pt\hbox{}
 }}
{\vbox {\centerline{Very Long Time Scales    }
\vskip10pt
\centerline{and Black Hole Thermal Equilibrium}
}}

\vskip0.4cm

\centerline{$\quad$ {\caps 
J.L.F. Barb\'on~$^{a,}$\foot{ On leave
from Departamento de F\'{\i}sica de Part\'{\i}culas da 
Universidade de Santiago de Compostela, Spain.} and 
E. Rabinovici~$^{b}$ 
}}
\vskip0.4cm

\centerline{{\sl $^a$ Theory Division, CERN, 
 CH-1211 Geneva 23, Switzerland}}
\centerline{{\tt 
barbon@cern.ch}} 

\vskip0.2cm

\centerline{{\sl $^b$ Racah Institute of Physics, The Hebrew University,  
 Jerusalem 91904, Israel}}
\centerline{{\tt eliezer@vms.huji.ac.il}}

\vskip0.4cm

\centerline{\bf ABSTRACT}

 \vskip 0.2cm
 
We estimate the very long time behaviour of correlation functions in 
the presence
of  eternal black holes. It was pointed out by Maldacena (hep-th 0106112)
that their vanishing would lead to a violation
of a unitarity-based bound. The value of the bound is obtained from the 
holographic dual field theory. The correlators indeed vanish in a
 semiclassical bulk approximation. 
We trace the origin of their
 vanishing to the continuum energy spectrum in the 
presence of event horizons. We elaborate on the  two very long time scales
 involved: 
one associated with the  black hole and the other with a 
thermal gas in the vacuum background.
 We find that assigning a role to the thermal gas  
background, as suggested in the above work, 
does restore the compliance with a time-averaged unitarity bound.
We also find that additional configurations are needed to explain the
expected time dependence of the Poincar\'e recurrences and their magnitude.
It is suggested that, while a semiclassical black hole does reproduce 
faithfully ``coarse grained'' properties of the system, additional 
dynamical features of the horizon may be necessary to resolve
a finer grained information-loss problem.     
In particular, an effectively formed stretched horizon could yield the desired
results.

\noindent

\vskip0.05cm

\Date{July 2003}
               

\vfill







\baselineskip=15pt

\newsec{Introduction}

\noindent

String theory seems a formidable multi-faceted fortress of consistency
 although it shows a
 less determined  side when needing to choose a vacuum or face Nature.
As large issues have failed so far to 
 breech its walls perhaps the nitty-gritty picking of small
details will suggest where it needs to be modified.
A context to study such important fine details is the question of possible information loss
 in black holes. 
An aspect of this problem is the search for a  more complete understanding of
 the propagation of strings on backgrounds 
with spacelike singularities. In fact one may need to await a nonperturbative
 formulation of string theory to resolve it. 
The AdS/CFT correspondence
offers such a formulation for some string backgrounds \refs\rads.
In particular it was found that field
theories
with a discrete energy spectrum are dual to strings propagating on  mixed
backgrounds sharing the same conformal
boundary. For example, for the case of an asymptotic ${\rm AdS}_5 \times
{\bf S}^5$
 both thermal AdS and a black hole in AdS need to be considered above a known 
temperature. 
The value of
 thermodynamical quantities such as the 
 entropy are reproduced and
accounted for by
the quarks and gluons of an $\CN=4$ supersymmetric
 CFT. This nonperturbative result gives, for the appropriate temperature, 
a classical black hole entropy: a standard field theory result encodes 
a less standard thermodynamical result in gravity.

These thermodynamical
 quantities measure gross
features of the system such as the total number of states. One would
like to probe the details further  and find to
what extent the low-energy supergravity ``master fields" arising
in the large-$N$ limit of the correspondence are sensitive also to the
 finer
structure of the system such as its energy level spacing. 
For example, on the nonperturbative field-theory side,
 the infrared  cutoff imposed by the finite
radius sphere ${\bf S}^3$ enforces a discrete energy spectrum. 
The bulk realization of this is less clear, yet, these finer
details can be related to the 
study of the possible unitary evolution of black holes. 
Nonperturbative completions of string
theory such as the above mentioned AdS/CFT correspondence leave no room
for unitarity violations as a matter of principle, but
the
detailed manner in which this is enforced in practice is yet to be
clarified. Indeed, here we search for the possible need of new dynamical effects.

Recently, a formal criterion for detecting the absence of information loss
has been proposed by Maldacena, based on the detailed study of thermal
equilibrium between a large eternal black hole and its thermal radiation in AdS
spacetime \refs\rmaldas. Since one studies a stationary problem involving
possibly  very large black holes, one has the impression that there are no
obstacles in obtaining definite answers within the semiclassical
approximation. However it turns out that this is actually a situation
in which the finer details are tested.
The suggestion in \refs{\rmaldas}
was to follow the detailed very long time
properties of correlation functions.    
In \refs{\rsuspouno, \rsuspodos, \rsuspotres} this program was pursued
as a particularly incisive probe of an analogous problem in quantum de Sitter
space.
It was suggested to examine the manner in which a form of a
time-independent lower bound on such correlation functions was respected.
Bounds of such nature are known to occur in systems subject to Poincar\'e
recurrences,   characteristic of finite bounded systems
which evolve in a unitary way, such as the nonperturbative 
holographic field theory ensures. In general one expects that an initial
perturbation of a given thermal system will be damped by the usual thermal
dissipation effects, this decay continues as
long as the time scale is too short to resolve possible gaps in the
spectrum. Eventually for a unitary system with a discrete spectrum the
perturbation is prevented from dying out. Moreover for any required
precision there will exist a time for which the correlation function 
returns to its initial value within the required precision. 

In \refs\rmaldas\ it was pointed out that semiclassically, on the 
bulk side, the correlation function does decay to zero 
in the presence of a black hole and thus violates 
the bound. 
 To the contrary, in a 
thermal AdS background
 the time-averaged correlation functions do respect a lower non-vanishing bound.
The proposal of \refs\rmaldas\ aims at recovering the recurrences by
summing over thermodynamically subleading backgrounds with discrete
spectrum, such as the thermal AdS manifold. This has the appealing feature of
requiring two spacetimes, with different topology, to contribute
 in the same string theory  
(for earlier work in this
 context see \refs{\rHPage, \rwithp, \rusex}.)   

In this note we discuss in more detail the nature of this proposal
(see \refs\rsolo\ for related discussions in
the case of three-dimensional black holes). In particular, we show that
Poincar\'e recurrences can indeed be absent for non-unitary systems.
 i.e. a system with non-coherent quantum evolution
\refs\rsdollar\ will show no recurrences despite having finite entropy.
On the other hand, we point out that the absence of Poincar\'e recurrences
in semiclassical correlation functions is tied to the existence of a
 continuous spectrum
of modes in the presence of black hole horizons. This is a property
of the ``bare" horizon, as described by General Relativity, and   
could be modified by nonperturbative  dynamical effects.
Thus  the absence of Poincar\'e
recurrences in this case may be  an artefact of the semiclassical
approximation.

We examine the proposal that recurrences are restored by 
topological diversity.
We present some evidence that, as tempting as the proposed resolution is,
it may not recapture some the expected detailed features of the system.
While it does lead to a result respecting  a lower bound on
 a time averaged normalized
correlation function, it actually seems to do it in a manner reflecting
neither the time scale of the recurrences nor the expected amplitude of
the correlation function. The time scale it provides is much shorter than
could be expected,  and the amplitude much smaller than could be expected,
 although their product does conspire so as to respect the bound. 

 This paper is
organized as follows. Sections 2 and 3 are a review of known facts that
serve to focus the problem. In Section 2 we review the intuition behind
the classical and   quantum bounds on correlation functions and their time
average. We stress that, for either unbounded or generic non-unitary
systems, the correlation functions and their averages vanish. In
Section 3 we recall the source of the continuous portion of the spectrum
of particles in the presence of a black hole and contrast it with the
discrete spectrum in the absence of event horizons.  In Section 4 we
evaluate the semiclassical prediction based on the topology change
processes suggested in \refs\rmaldas. We estimate the time-averaged lower
bounds in two examples: the Schwarzchild black hole in a box and AdS black
holes in various dimensions. Some particular  behaviour is noted
for the three dimensional case. 

In a sense we claim that while the semiclassical black hole master field
picture is very valuable for computing inclusive properties of the system
such as entropy and free energy, the appreciation of finer properties such
as the Poincar\'e recurrences may require finer tools. We end by
discussing the possible form of such tools.

\newsec{Unitarity Versus Poincar\'e Recurrences}

\noindent

Poincar\'e recurrences in classical systems  
are a consequence of
the compactness of the energy shell in phase space, together
with Liouville's theorem, which states that the canonical flow preserves
phase-space volume. It then follows that the time evolution of any
finite-volume 
domain of phase space, $\Omega$,  will eventually intersect  itself.
To see this, consider a series of discrete ``snapshots" of the time
evolution with fixed time increment
$ \Omega, \; U(\Omega), \; U^2 (\Omega), \; \dots\;, U^n (\Omega), \dots$ 
Since all the images of $\Omega$ have the same volume and the phase space
is compact, there must be  two images in the list with  
a non-vanishing intersection:
$U^k (\Omega) \cap U^l (\Omega) \neq \emptyset$. It follows that
$\Omega \cap U^n (\Omega) \neq \emptyset$ for $n=k-l$.
This is Poincar\'e's ``eternal return" \refs\rarnold. 
There are two  independent ways in which we may
 evade such eternal return. Either we 
may have access to a
{\it non compact} phase space, or we may consider time flows
that do not preserve phase-space volume.

The quantum analogue of Liouville's theorem is the
unitarity of the time evolution. The compactness of phase space corresponds
to having a finite number of states under the energy shell, i.e. the
relevant energy spectrum is {\it discrete}. 
The phase-space volume
of the initial set $\Omega$ is roughly related to the degree of ``purity"
of the initial quantum state.  

A useful quantity to monitor quantum recurrences is the time correlation
function of an observable $A$, 
\eqn\timeco{
G_\rho \,(t) = \Tr \,\left[\;\rho \; A(t) \,A(0) \right]
\,,} 
where we take $t>0$ and $\rho$ denotes the density matrix characterizing
the initial
 state of the system. The decay of this correlation function in time gives
information about the dissipation of perturbations of typical equilibrium states,
such as the canonical thermal state at
temperature $1/\beta$: 
\eqn\therm{
\rho_\beta =  {e^{-\beta H} \over Z(\beta)},
}
where $Z(\beta) =  
 \Tr \exp(-\beta H)$ is the canonical partition function. A more fundamental 
equilibrium state for bounded isolated systems is actually the
microcanonical state
\eqn\microc{
\rho_E = e^{-S(E)}\;\Theta(E-H) \;,
}
where $\Theta$ is the  step function and 
\eqn\micen{
S(E) = \log \Tr \Theta(E-H)
}
is the microcanonical entropy.  
For a bounded system with discrete energy spectrum and unitary evolution, 
\eqn\unitt{
A(t) = e^{itH} \;A(0)\; e^{-itH}
,}
the thermal dissipation cannot be complete at very long times. For example, the
 microcanonical correlation function takes the form  
\eqn\bounc{
G_E\,(t) = e^{- S(E)}\; \sum_{E_i, E_j \leq E} |A_{ij}|^2 \;
e^{i(E_i -E_j)t}
\;.}
 In terms of the frequencies $\omega_{ij} = E_i -E_j$, we can define
 the so-called Heisenberg time (c.f. \refs\rsrednicki) by $t_H = 1/\bra \omega 
 \ket$, where $\bra \omega \ket$ denotes an average of the frequencies that
 we may estimate as the total energy divided by the total number of states:
 $\bra \omega \ket \sim E \,\exp(-S(E))$. The Heisenberg time is the time
 scale that reflects the discreteness of the spectrum, i.e. for $t \ll t_H$
 we can
approximate the spectrum as continuous. In this case, 
if the matrix elements of $A$ in the
energy basis have a frequency width  $\Gamma$, the correlator will decay
with a characteristic lifetime of order $\Gamma^{-1}$, i.e. we expect standard
dissipative behaviour for times $\Gamma^{-1} \ll t \ll t_H$.

For $t>t_H$ most of the phases in \bounc\ would have completed a period
and the function $G_E \,(t)$ starts to show large irregular   oscillations.   
 In fact, $G_E \,(t)$ as defined by \bounc\
is a quasiperiodic function of time;  
 despite
 thermal damping,   it  
returns arbitrarily close to the initial value  over
periods of the order of the recurrence time. In order to estimate this 
recurrence time, let us assume that we have $\CN$ different  
frequencies $\omega_{ij} = E_i - E_j$ that are rationally independent.
 Then we can picture the  phases  
as a set of $\CN$ clocks running at angular velocities $\omega_{ij}
$ (c.f. \refs\rthirring).
 If we demand that they return within a given angular accuracy $\Delta \alpha$,
the measure of this configuration is  $(\Delta \alpha /2\pi)^\CN$, and the  
recurrence time is of order $(2\pi/ \Delta \alpha )^\CN /\bra \omega \ket$,
where $\bra \omega\ket$ is the average frequency. Since $\CN \sim \exp (2S)$, 
 the recurrence or Poincar\'e time scales as a double exponential
in the entropy. These scaling properties of the Heisenberg and Poincar\'e time
scales have been explicitly tested in a numerical model in Ref. \refs\rsuspouno.   

Notice that setting 
  $\Delta \alpha \sim 2\pi$ we obtain the Heisenberg time scale, corresponding
  to $\CO(1)$ ``resurgences" with no emphasis on accuracy, i.e. the Heisenberg
time is the smallest possible Poincar\'e time.  
If the spectrum shows special regularities the                  
recurrence time can be smaller.
For example,  all energy levels of a $1+1$ dimensional free quantum field
in a circle of radius  
 $R$  are multiples of a single frequency, in which case
the correlation function is strictly periodic with period $2\pi R$.

For large systems with positive specific heat  the same results should follow
for
the canonical state $\rho_\beta$, in terms of the canonical entropy and internal
energy. In this case the energy sums in the analogue  of \bounc\ are not bounded by the
total energy, but only damped by the Boltzman factor $\exp(-\beta E)$. In general, it
may be necessary to bound the high-energy off-diagonal matrix elements $|A_{ij}|^2$ in
order to ensure convergence of the canonical
correlation function $G_\beta \,(t)$, particularly in
the limit $t\rightarrow 0$. In a local quantum field
theory this may require paying due attention to $i\epsilon$ prescriptions, and
 perhaps working with smeared or regularized operators. 

 We conclude that time scales associated to large-scale fluctuations or 
Poincar\'e recurrences depend exponentially on the entropy. Systems with continuous
spectrum have strictly infinite total entropy, and both Heisenberg and Poincar\'e
times are infinite. 

\subsec{Non-Unitary Evolution}

\noindent

To further sharpen
the relation between recurrences and unitarity, let us consider
the loss of quantum coherence  envisaged by Hawking in \refs\rsdollar.
 Time evolutions of this sort violate quantum coherence but preserve
hermiticity, positivity and normalization of the density matrix. It was
shown in \refs\rpesk\ that these systems can be
generically modelled by ordinary quantum mechanics coupled to a
Gaussian random noise.

As a simple example, let us consider  a perturbation of the
 Hamiltonian  by an
operator of the form
\eqn\pertt{
H \longrightarrow H + C\,J(t)\;,}
where $C$ is a fixed operator that commutes with $H$ but not with  the
observable $A$,  and $J(t)$ is a gaussian random noise with
covariance
\eqn\cob{
\bra \,J(t)\,J(t') \,\ket_{J} = h \;\delta(t-t')\;.}
We also assume that the operator $C$ is non-degenerate in the energy basis.
Then, substituting  the formula \unitt\ by
\eqn\uss{
A(t) = e^{itH + iC\int_0^t dt' J(t')} \; A(0) \;e^{-itH -iC\int_0^t dt' J(t')}
}
we obtain, for the general correlator  \timeco,  
\eqn\nois{
G_\rho\,(t)_J =  \sum_{i,j,k} \rho_{ij} \;e^{i(E_j - E_k)\,t} \;A_{jk} \;A_{ki} \;
\; \left\langle \,e^{i(C_j - C_k)\,\int_0^t dt' J(t')} \right\rangle_{J}
\;,}
where $C_i$ are the eigenvalues of $C$ in the energy basis.  
  Evaluating the Gaussian average
one obtains
\eqn\nnni{
\int \prod_{t'} \left[{dJ(t') \over \sqrt{2\pi}}\right] \;e^{-\int dt' |J(t')|^2 /2h}
\; e^{i(C_j - C_k)\,\int_0^t dt' J(t')} =
 e^{-h \, |C_j - C_k|^2 \,t /2}
 \;,}
a damping factor that eliminates all possible recurrences as $t\rightarrow
\infty$. 
This result implies that Poincar\'e recurrences
may be used as a criterion for unitarity, provided the energy spectrum
is truly discrete. On the other hand, one may have a perfectly unitary
evolution, and still miss the Poincar\'e recurrences because the
spectrum is continuous.

\subsec{Time Averages}

\noindent

 We now return to unitary time evolutions and define a quantitative measure of the
quasiperiodicity of correlators. First we define a normalized correlator
\eqn\rat{ 
L(t) \equiv \left|{G(t) \over G(0)}\right|^2
\;,}
which is independent of the overall normalization of the operators.  Then we can
measure the long time resurgences of $L(t)$ by the value of   the time average
\eqn\ave{
\bL   \equiv \lim_{T\to \infty} {1\over
T} \int_0^T dt \;L(t) \; . 
}
A rather general estimate of the quantity $\bL$ can be obtained for
  correlations in the canonical ensemble  
\eqn\genco{
G_\beta \,(t) = \Tr\left[ \,\rho_\beta\,A(t) \,B(0)\,\right]
={1\over Z(\beta)} \sum_{i,j} e^{-\beta E_i} A_{ij} \;B_{ji} \;e^{i(E_i -E_j)t}
\,,}
for general observables $A,B$  with vanishing diagonal matrix elements
in the basis of energy eigenvectors. Following \refs\rsuspodos\ we have  
\eqn\ratio{
\bL = {\sum_{i,j,k,l} e^{-\beta(E_i + E_k)}
\, A_{ij} \; B_{ij}^* \; A_{kl}^* \; B_{kl} \;
\delta_{i - j - k +l}
 \over
 \sum_{i, j,k,l} e^{-\beta(E_i + E_k)}
 \,  A_{ij} \; B_{ij}^* \; A_{kl}^* \; B_{kl} }
 \,.}
The main
difference between numerator and denominator is the Kronecker delta 
that eliminates one of the index sums. Since all terms in the numerator
and denominator are essentially identical, we can imagine that
the denominator is bigger roughly  by a factor of
 the number of states that effectively
 contribute, i.e. the thermodynamical degeneracy $\exp(S)$. 
A more formal argument can be given for large systems,
 by passing to continuum notation, approximating
the sums over  discrete energy levels by integrals according to the rule:  
\eqn\cont{
\sum_n f(E_n) \approx \int dn \,f(E(n))= \int d\mu(E)\,f(E)\;,
}
where 
\eqn\meas{
d\mu(E) = dE\,{dn(E) \over dE}\;.} 
Using $n(E) = \exp S(E)$ and defining the microcanonical temperature
function
$
\beta(E) = dS(E) / dE $ we have
\eqn\contt{
d\mu(E) = dE\,\beta(E)\,e^{S(E)}
\;.} 
In \contt, $S(E)$  is a smooth function of the continuous energy variable $E$. It  
has the  interpretation of an interpolating function that agrees with \micen\ on the
discrete energy levels. This {\it does not} mean that the spectrum is continuous and, 
in particular, the total entropy $S(E)$ is a finite function.
 
The time average \ratio\ can be approximated as    
\eqn\cono{
\bL \approx {\int d\mu(E_1)\; 
d\mu(E_2)\; d\mu(E_3)  \;A_{E_1, E_2} \,B_{E_1, E_2}^* \,A_{E_3, E'
}^* \,B_{E_3, E'} \;e^{-\beta(E_1 + E_3)}
\over
\int d\mu(E_1)
\;d\mu(E_2)\; d\mu(E_3) \;d\mu(E_4)
 \; A_{E_1, E_2} \,B_{E_1, E_2}^* \,A_{E_3, E_4
 }^* \,B_{E_3, E_4}
  \;e^{-\beta(E_1 + E_3)}
  }}
  where $E' = E_2 + E_3 -E_1$.
  We can now evaluate the integrals over $E_1$ and $E_3$ in the saddle
  point approximation. Both saddle points are identical and located at
  $E=E(\beta) =E_\beta$, the internal energy that solves the equation:
  \eqn\microc{
  \left({\partial S \over \partial E} \right)_{E=E_\beta} = \beta
  \,.}
  For large systems with positive specific heat $C_V \gg 1$ we may  neglect
the fluctuations around   the saddle point and get
  \eqn\appp{
  \bL \approx {\int
  dE_2 \,\beta(E_2)  \;|A(E_\beta, E_2)|^2 \;|B(E_\beta, E_2)|^2
   \;e^{S(E_2) }
   \over
   \left|\int dE_2 \,\beta( E_2) 
    \;A(E_\beta, E_2) \;B(E_\beta, E_2)^*
     \;e^{ S(E_2) }\right|^2
     }\;.}
     Notice that the factors proportional to the partition function
      at the saddle point, 
      $\exp\left(S(\beta)-\beta E(\beta)\right)$, 
       cancel between numerator and denominator. Now
   let us assume that the operators $A$ and $B$
   have matrix elements around $E(\beta)$
  consisting of a band of width $\Gamma /2
 \ll E(\beta)$ and  approximate
  the integrands  by their mean value within this band\foot{In local quantum field theory,
this technical requirement may depend on an appropriate regularization of the operators,
 particularly in the definition of $G_\beta\,(0)$.}. Then 
  we obtain the estimate (c.f. \refs\rsuspodos)  
 \eqn\fin{ 
\bL \sim {\beta \,\Gamma  \cdot
 e^{S(
 \beta)} \over \left[\beta\,\Gamma \cdot 
 e^{ S(\beta)}\right]^2} \sim {e^{-S(\beta)} 
 \over \beta\,\Gamma} 
 \;,}
up to a ratio of matrix elements corresponding to energies close to $E_\beta$. The
proportionality factor in \fin\ can also get contributions that depend on
the precise definition of entropy function. For example, replacing the step function
in \microc\ and \micen\ by a Kronecker delta has the effect of replacing $\beta$
by $E_\beta$ in \fin. Apart from these logarithmic ambiguities in its definition, the  
entropy factor in the exponent has corrections of $\CO(\beta \,\Gamma)$, a quantity
that we assume finite in the thermodynamical or semiclassical limits.  

Thus, $\bL$ is bounded strictly above zero provided the thermodynamic entropy
is finite. This shows  that a  
continuous spectrum  (such
as that of an unbounded system) implies $\bL=0$ and no Poincar\'e recurrences.

One can derive analogous bounds for different types of correlators. One
interesting case is the correlation between operator insertions on 
disjoint thermo-field doubles (c.f. \refs\rmaldas). In this formalism one
works with pure states in a doubled Hilbert space, with a precise entanglement
between the two copies. Standard thermal correlators correspond to 
operator insertions on a single copy. Operator insertions on different copies
can be interpreted as measurements of slightly non-thermal states.  The
relevant correlation functions can be obtained by analytic continuation.
For example, for the two-point function on different copies, 
\eqn\lrt{
G_\beta\,(t)_{\rm LR} \equiv \Tr\,\left[\,\rho_\beta\, A(t-i\beta/2)\,B(0)\,\right]\;.}
Aplying similar methods to this correlator\foot{One interesting difference with
\genco\ is that now all energy sums are damped by Boltzman factors, improving the
convergence in the $t\rightarrow 0$ limit.}
   we obtain the estimate of the
appropriate time average:
\eqn\ett{
\bL_{\rm LR} \sim \exp(-K(\beta))\,,\;\;{\rm where}\;
\qquad K(\beta) = 2\beta\,\left(F(\beta)-
F(\beta/2)\right) - S(\beta)\;, 
}
and $F(\beta) = E(\beta)-S(\beta)/\beta$ denotes the canonical  
free energy.

\newsec{No Recurrences in the Presence of Event  Horizons}

\noindent

For a bounded gravitating system, black holes should dominate the thermal ensemble
at high energies. In that case we expect the bound \fin\ to hold with $\beta$
the inverse Hawking temperature and 
$S(\beta) = A_H /4G_{\rm N}$ the Bekenstein--Hawking
entropy. Thus, the expected recurrence index  for black holes
has a nonperturbative scaling with Newton's constant: $\bL \sim \exp(-1/G_{\rm N})$. 
This suggests that it might be calculable in the semiclassical approximation.  

On the other hand, the infinite redshift of an event horizon  implies that a 
 quantum field  can accumulate
an infinite number of modes there, with a finite cost of energy (c.f. \refs\rbrick).
Hence, perturbative correlation functions should reflect a continuous spectrum of
excitations and produce  $\bL =0$.
 In this section we review this known fact and identify some
related subtleties of the Euclidean formalism and the procedures of 
analytic continuation.

\subsec{The Continuous Spectrum}

\noindent

Let us consider a static background metric  of the general
 form
\eqn\bhk{
ds^2 = -g(r)\,dt^2 + {dr^2 \over g(r)} + r^2 \,\d\Omega_{d-2}^2 \;,}
where $r=r_0$ is a non-degenerate event horizon, $g(r_0)=0$, with
Hawking temperature $\beta^{-1} = g'(r_0)/4\pi$.

The natural analogue of \genco\ in the background of a black hole is
a thermal correlation function of the form
\eqn\nueve{
\CG (t, t') = {\Tr \left[\,
e^{-\beta \CH}\;\phi(t,r,  \Omega)
\, \phi(t',r', \Omega') \,\right] \over \Tr \,e^{-\beta\CH} }\;,}
where $t-t' >0$ and $\phi$ is a perturbative field in the background \bhk. In the  
notation of Section 2, this would correspond to $A(t) = \phi (t, r, \Omega)$ and
$B(t') = \phi(t', r', \Omega')$. The splitting of spatial coordinates $r, \Omega$
avoids ultraviolet divergences in the $t-t' \rightarrow 0$ limit, although there
remain possible on-shell singularities that must be handled with the appropriate $i\epsilon$
prescription. 

 The operator
 $\CH$ in \nueve\ is the Hamiltonian with respect to the asymptotic time variable
$t$, and is
defined {\it perturbatively} around \bhk.  In the free
 approximation for a bosonic  field $\phi$ we have
\eqn\vein{
\CH_{\rm free}= \sum_{\omega} \omega \,\left( N_\omega  
 + \shalf \right)
\;,}
where $N_\omega$ is the occupation number of the mode with frequency $\omega$.
The large-time behaviour of such correlation functions reduces then
to the properties of the frequency spectrum $\omega$.

In the free approximation, we can obtain the frequency spectrum by studying
the  wave equation 
\eqn\dsei{
(\nabla^2 - m^2) \,\phi (x) =0
\,}
in a basis of modes with definite frequency  
\eqn\bmod{
\phi(t,r,\Omega)  = {1 \over \sqrt{2\omega}} \; e^{-i\omega t} \;
\phi_\omega \,(r,\Omega)\;,}
and normalized in the  Klein--Gordon inner product:
\eqn\kg{
\bra\,\psi\,|\, \varphi\,\ket_{\rm KG} = i \int_{t={\rm const}} d\sigma^\mu \;
\psi^* \,\darr{\pt}_\mu\, \varphi\;.}
 Then, the free approximation to the thermal Green's function
takes the usual form
\eqn\gree{
\CG(t, t')_{\rm free} =
\sum_{\omega} {\phi_{\omega} (r,\Omega) \,\phi_{\omega} (r',\Omega')^*
\over 2\omega} \left[\;(1+n_\omega)\, e^{-i\omega\, (t-t')} +
n_\omega \,e^{i\omega\, (t-t')} \;\right]
\;,}
with
$
n_\omega = \left(e^{\beta\omega} -1\right)^{-1}
$ the average occupation number in the thermal ensemble.
Replacing $t\rightarrow t-i\beta/2$ in \gree\ 
we obtain the analogue of \lrt, which is ultraviolet-finite 
in the $t-t' \rightarrow 0$ limit, even for $r=r'$, $\Omega=\Omega'$,
 and is free from on-shell singularities. However, its long
time behaviour is governed by the same frequency spectrum $\omega$.

 We can further factor out the $SO(d-1)$ symmetry by writing
\eqn\doch{
 \phi_{\omega} (r, \Omega)
   = r^{2-d \over 2}
   \;f_{\omega,\ell} (r) \; Y_\ell (\Omega)
   \,,}
   with $Y_\ell$ the appropriate spherical harmonic on ${\bf S}^{d-2}$:
   $$
   -\nabla^2_{\Omega} \,Y_\ell (\Omega) = C_\ell \,Y_\ell (\Omega)\,.
   $$
   The frequency and radial wave function are determined by the
Schr\"odinger problem
   \eqn\veun{
 \left[- {d^2 \over dr_*^2} +
 V_{\rm eff} (r_*) \right] \,f_{\omega,\ell} (r_*)
   = \omega^2  \,f_{\omega,\ell} (r_*)\;,}
   with
   \eqn\vedo{
   V_{\rm eff} = {d-2 \over 2}\,g(r)\left({g'(r)\over r}  +
   {d-4 \over 2r^2} \,g(r) \right) + g(r)  \left({C_\ell
   \over r^2} + m^2 \right)
   \;.}
   Here we have defined  the
 Regge--Wheeler or   ``tortoise" coordinate
   $ dr_* = dr / g(r)
  $. In terms of the $r_*$ coordinate, the eigenvalue problem \veun\ inherits
a standard $L^2$ inner product from \kg. 

For asymptotically Minkowskian spacetimes, $g(r) \rightarrow 1$ as $r\rightarrow \infty$,
the effective potential
asymptotes to $V_{\rm eff} \rightarrow m^2 $, and we obtain
the standard continuum spectrum of frequencies for $\omega>m$.  Near a non-degenerate
horizon, $g(r_0) =0$, the effective potential shows the universal scaling
\eqn\univh{
 V_{\rm eff} (r_*) \;\propto \;e^{4\pi r_* /\beta}\;, \;\;\;\;{\rm as}\;\;\;r_*
\rightarrow -\infty\;,
  }
   with $1/\beta
= g'(r_0) /4\pi$ the Hawking temperature.
     We see that the spectrum of {\it positive real}
frequencies will be continuous, quite independently of the properties of
the potential for $r\gg r_0$. In particular, enclosing the system in a large
box by placing a reflecting
 wall at large and finite $r$ fails to discretize the spectrum,
due to the exponential tail \univh.

It is instructive here
to consider the large $r$
 Anti-de Sitter   asymptotics $g(r) \rightarrow r^2 /R^2$ from the
point of view of the effective Schr\"odinger problem in tortoise coordinates. For
$r\rightarrow \infty$ the tortoise coordinate approaches a maximum value
$(r_*)_{\rm max} = \pi R/2$. In addition, the effective potential diverges
as $r_* \rightarrow \pi R/2$.
  So, we recover the known fact that AdS spaces behave  effectively as finite-volume
  cavities.
 For vacuum AdS in global coordinates, $g(r) = 1+ r^2 /R^2$,
 the tortoise coordinate
    near the origin is $r_*\sim r$,
     so that  the range of the effective potential
     is finite: $r_*\in [0,\pi R/2]$. The
 spectrum is thus discrete in pure AdS, and continuous in the AdS black hole.
However, the Poincar\'e patch of vacuum AdS, with $g(r) = r^2 /R^2$ has
continuous spectrum, as $r=0$ is a degenerate horizon with $g(0) = g'(0) =0$.

The frequency  spectrum in the presence of horizons is thus continuous in the
free approximation. Perturbative corrections in the black hole background
 are constructed in terms of the
free Green's functions, and will  not discretize the spectrum (unless
perturbation theory itself breaks down.)
Hence, we have
perturbative correlation functions $\CG(t)$ with infinite Heisenberg time.
Defining the corresponding time average in the background of interest,
\eqn\timeav{
{\overline \CL}
 \equiv \lim_{T\to\infty} {1\over T} \int_0^T dt \left|{\CG(t) \over \CG(0)}
\right|^2
\;,}
we will find ${\overline \CL}=0$
 in the presence of static horizons, at least within
the perturbative formalism sketched here.
Since
$ t-t' >0$ in our Green's functions, they are actually
 retarded correlators,  and the large-time decay
can be understood in terms of complex frequencies, a thermal
analog  of
the so-called ``quasinormal modes" \refs\rsolo.

\subsec{Euclidean Formalism}

\noindent

The vanishing of the recurrence index, $\bL=0$, in the previous discussion is
tied to the continuous spectrum of the perturbative Hamiltonian $\CH = i\,\pt/\pt t$. 
In the same approximation, the perturbative partition function $\Tr \exp(-\beta \CH)$
that appears in the normalization of \nueve\ is divergent. This means that  the 
 defining expression \nueve\ is somewhat formal (although \gree\ does make sense
even for continuous spectrum.) A more rigorous
 definition of thermal correlation
functions is by analytic continuation from Euclidean correlation functions. This
formalism also offers a vantage point of view on  
 the issues of regularization and renormalization of ultraviolet
divergences.    

In the semiclassical approach to thermal effects in quantum gravity,
one starts from smooth Euclidean backgrounds with appropriate
boundary conditions, c.f. \refs\rHH. For the case of black holes
in canonical ensembles, one encounters 
 Euclidean manifolds $X$, with metrics of the form
\eqn\eucl{
ds^2 = g(r)\,d\tau^2 + {dr^2 \over g(r) } + r^2 \,d\Omega_{d-2}^2\;,}
with $\tau \equiv \tau + \beta$ and $\beta$ the
inverse Hawking temperature. With
this periodicity of the Euclidean time $\tau$ the metric \eucl\ is smooth at
the horizon, defined by $g(r_0)=0$,
and\foot{Everything we say in this section can be generalized to the more general
case of metrics $ds^2 = g(r)\,d\tau^2 + f(r)^{-1} \,dr^2 + r^2 ds^2_K$, with
$f(r_0) = g(r_0) =0$ and $K$ a $(d-2)$-dimensional  compact manifold.}
  is restricted to the region $r\geq r_0$.

The thermal partition function is given by $Z(\beta) = \exp(-I(X))$, where $I(X)$
is the Euclidean effective action evaluated on $X$. As a function of $\beta$, it is
 interpreted as the   canonical free energy, 
and determines all thermodynamic functions,
including the  correct black-hole entropy \refs\rGH: 
\eqn\cbh{
S(X) = (\beta\,\partial_\beta -1)\,I(X)\;.}
We shall assume that the asymptotic boundary conditions ensure the stability of the
canonical ensemble, in the sense that the specific heat is positive: $C_V >0$. For
black hole states this will require working on a finite box, or in AdS space. 

 In the gravity
 perturbative expansion, the Euclidean effective action  
takes the form
\eqn\perts{
I(X) = \sum_{n=0}^{\infty} \lambda^{2n-2} \;I_n (X)\;,}
where the $I_n$ are functions of $\beta$ and whatever other moduli we can
adscribe to $X$. 
  The effective expansion parameter is given by the
string coupling $\lambda = g_s$ or some multiple of Newton's constant
$\lambda^2 = G_{\rm N} /\ell_{\rm eff}^{d-2}$
 in low-energy descriptions with effective
cutoff at length scale $\ell_{\rm eff}$.
Notice that the leading  term $\lambda^{-2} I_0 = I_{cl}$
 is precisely the {\it classical}
approximation to the effective action and is of $\CO(1/G_{\rm N})$.
 Although these manipulations are usually
carried out within a low-energy field theory, Euclidean backgrounds
defining worldsheet conformal field theories are also the starting point
of perturbative string theory. Even  at a nonperturbative level,
the Euclidean approach to gravitational
thermodynamics has been shown to integrate nicely within the AdS/CFT
correspondence \refs\rwithp.

Following \refs{\rHH, \rgibperry}\ we can define real-time 
 properties of the thermal ensemble  by
standard analytic continuation to Lorentzian metrics: $\tau = i\,t$. For
instance, Euclidean correlation functions of local fields in \eucl\ 
can be used to derive Lorentzian counterparts  on the different patches
of the extended Lorentzian manifold, such as \nueve, or 
the vacuum Green's function on  the full Kruskal extension
of the Schwarzschild metric.  

A formal path integral formula for Euclidean Green's functions is
\eqn\siete{
\CG^E (\tau, \tau') =
{\int d\mu (\phi, \dots)
\;\phi(\tau, r,  \Omega) \, \phi(\tau',r', \Omega')
\;e^{-\half \int_0^\beta  (\pt \phi)^2 +
m^2 \phi^2 + \dots}
\over \int d\mu (\phi, \dots)
\;e^{-\half \int_0^\beta  (\pt \phi)^2 +m^2 \phi^2
+ \dots}
}}
where
 the dots represent the contribution of other fields
 to the measure and the action, together with
 the perturbative interactions. In principle, \siete\ is defined
in a low-energy field theory, but we could imagine that
such correlation functions exist in
an off-shell definition of string field
theory.  In this expression,
 $d\mu(\phi, \dots)$
is the formal  measure over the field $\phi$, together
with the metric fluctuations
and all the other bulk degrees of freedom with canonical
thermal boundary conditions. 

A more rigorous definition is possible in 
 the context of the AdS/CFT correspondence,
where one can define correlators of local CFT operators by 
  taking the $r\rightarrow \infty$ limit
in the bulk field insertions, with wave-function factors $(r/R)^\Delta$
for fields dual to a local CFT operator of conformal dimension $\Delta$. 
Correlation functions such as \siete, in terms of local operators in the
bulk, can be interpreted as correlators of spatially nonlocal operators in the  CFT. 
Since our discussion of recurrences in Section 2  only requires the operators
to be local in time,  
  we may carry our arguments directly in terms of \siete.  

The Euclidean time direction in \eucl\ is isometric, and $\CG^E (\tau, \tau')$
is only a function of $\Delta \tau = \tau -\tau'$. 
After $\CG^E (\tau, \tau')$ has been renormalized on the Euclidean manifold $X$, we can
define \nueve\ by  standard analytic continuation in the time argument, $-i\Delta \tau  =
\Delta t =t-t' >0$, a procedure that takes care of the appropriate $i\epsilon$ 
prescription. From the same Hartle--Hawking Green's function 
one can also define   correlators on different thermo-field doubles, 
 which do not suffer
from on-shell singularities,  by the prescription \lrt.   

This definition makes it 
somewhat paradoxical that the Lorentzian Green's function $\CG(t,t')$ should
show a continuous spectrum of excitations localized at the horizon, because
$r=r_0$ is a perfectly regular point of the manifold $X$. 
 The Euclidean Green's function can be
written as
\eqn\ive{
\CG^E (\tau, \tau') = \bra \,\tau\,|\,(K_X)^{-1}\,|\,\tau'\,\ket + \dots\;,}
where the dots stand for interaction 
corrections and $K_X$ is the kinetic operator on $X$. For
a free scalar, it is given by $K_X = -\nabla_X^2 + m^2$. 
The manifold $X$ is smooth at $r=r_0$ and
the spectrum of $K_X$ becomes discrete once we impose a large-volume cutoff.
 Then, how is
it possible for \gree\ to have a spectral decomposition with continuous frequency
spectrum? 
The ``disease"  of the continuous
 spectrum has been tied to the phenomenon of divergences in
the calculation of black hole entropy  \refs\rbrick\  and the question of
their renormalization \refs\rsusug. However, here
we are obtaining a consequence of this ``disease" at the level of a manifestly 
ultraviolet-finite quantity,
such as the long-time free propagator. In this sense,   
an {\it infrared} interpretation of the   
 divergence in \refs\rbrick\ seems more natural,
 along the lines of \refs{\rhordiv, \rroberto}.     

This phenomenon     
 can be traced back to a subtlety in the analytic continuation from
\siete\ to \nueve, which in turn originates on the peculiar topology of the
Euclidean black hole manifold $X$.
 We sketch here the main points, referring the reader to
\refs\rroberto\ for more details. 

 In order to write a Hamiltonian representation with respect
to $\CH = i \,\partial/\partial t$
 one must foliate the spacetime in $t={\rm constant}$
hypersurfaces. In the original Euclidean manifold $X$ this corresponds to $\tau =
{\rm constant}$ hypersurfaces. 
However, the Euclidean time orbits are contractible on $X$, and the
 horizon $r=r_0$ is a singularity of the folliation,
because it is invariant under translations of $\tau$. 
 For this reason we can expect some subtle behaviour
of the Hamiltonian representation \nueve. One way of rendering the folliation non-singular
is to remove the submanifold at the  point 
 $r=r_0$ and define the correlation functions by continuity.
Then the manifold $X' \equiv X-\{r=r_0 \}$
 has a smooth time folliation and differs from $X$
by a set of measure zero. However, it has different topology from $X$ and the ``boundary"
at $r=r_0$ is non-compact.

We can now consider the operator 
\eqn\opp{
{\widetilde K}_{X'} = |g_{00}|\,K_X \;,}
which is well defined on $X'$ and has a very simple time dependence. For the
metrics of the form \eucl,  
\eqn\expli{
{\widetilde K}_{X'}
= -\pt_\tau^2 - g(r) \,r^{2-d} \,\pt_r \; r^{d-2} \;\pt_r - g(r)\;\nabla_\Omega^2 +m^2\;.}
 Instead of the standard covariant inner product 
\eqn\inn{
\bra \,\psi \,|\,\varphi\,\ket = \int_X  d^d x \;\sqrt{|g|}\,\psi^* (x) \,\varphi(x)}
we may define a rescaled inner product on $X'$:
\eqn\resc{
\bra\!\bra \,\psi \,|\,\varphi\,\ket\!\ket = \int_{X'}  d^d x \; \mu(x) 
\,\psi^* (x) \,\varphi(x)\;, \qquad {\rm where}\;\;\;\;\mu(x) \equiv {\sqrt{
|g|} \over |g_{00}|}\;.}
Then, the Green's function
\eqn\ott{
{\widetilde \CG}^E (\tau, \tau') = 
\bra\!\bra \,\tau\,|\,({\widetilde K}_{X'})^{-1}\, | \,\tau' \,
\ket\!\ket } 
satisfies the same differential equation as the Hartle--Hawking Green's function \ive, when
restricted to $X'$. On the other hand, the eigenvalue problem for the
operator  ${\widetilde K}_{X'}$
is given by
\eqn\eigv{
{\widetilde K}_{X'} \;{\widetilde \psi}_{n,\omega} (\tau, r, \Omega) = \lambda_{n,\omega}
\;{\widetilde \psi}_{n,\omega} (\tau, r, \Omega)\;,}
with 
\eqn\sole{
\lambda_{n,\omega} = {4\pi^2 n^2 \over \beta^2} + \omega^2 \;,\qquad 
{\widetilde \psi}_{n,\omega} (\tau, r, \Omega)= {1\over \sqrt{\beta}} \;e^{2\pi i \tau
/\beta} \;\phi_\omega (r,\Omega)\;,} 
  where $n\in {\bf Z}$ and $\omega^2$ solves
\veun. Therefore, on $X'$ we can write
\eqn\onx{
{\widetilde \CG}^E (\tau, \tau') = {1\over \beta} 
\sum_n \sum_\omega {e^{2\pi i n(\tau-\tau')/\beta} \,
\phi_\omega (r,\Omega) \,\phi^*_\omega (r', \Omega') \over 
\omega^2 + 4\pi^2 n^2 / \beta^2
}\;.}
Evaluating now the sum over $n$ using the identity
\eqn\ded{
{1\over \beta} \sum_{n\in {\bf Z}} {e^{2\pi i n \Delta \tau / \beta} \over
{4\pi^2 n^2 \over \beta^2} + \omega^2 } = {1 \over 2 \omega} {
\cosh \omega (\Delta \tau - \beta/2) \over \sinh (\beta \omega /2)}}
 and performing the analytic continuation $\Delta \tau = i\,\Delta t$ we obtain
\gree.

 Since \ive\ and \onx\ satisfy the same differential equation, we expect that
${\widetilde \CG}^E$   gives exactly the Hartle--Hawking Green's function on $X$, when
extended by continuity from $X'$. This last step is not completely free of ambiguities,
since the operator ${\widetilde K}_{X'}$  itself is not well defined at $r=r_0$.  However,
this question can be settled by an explicit calculation in the near-horizon approximation.

\subsec{The Near-Horizon Limit}

\noindent

The emergence of 
continuous spectrum out of smooth Euclidean Green's functions can be 
 illustrated  with the explicit example of
perturbations in the near-horizon region. For $r\approx r_0$ we can 
approximate the metric by Rindler space, with a pure exponential potential
in   
 \univh. Setting $\xi = 2 \sqrt{(r-r_0)/g'(r_0)}$ and
$\theta = 2\pi \tau /\beta$, the Euclidean Rindler space in the 
vicinity of the horizon $r\approx r_0$ becomes a
copy of flat ${\bf R}^d$ space:
\eqn\rin{
ds^2 \approx \xi^2 \,d\theta^2 + d\xi^2 + d{\bf y}^{\;2}\,.}  
 In Lorentzian signature, the wave functions on
the exponential potential are given by Bessel functions
\eqn\bess{
\phi_{{\bf p},\,\omega} (\xi, {\bf y}) = {e^{i{\bf p}\cdot {\bf y}} 
\over \pi\,(2\pi)^{d-2 \over 2}}
 \;\sqrt{2\nu \, \sinh(\pi \nu)}\;
K_{i\nu} (\mu\,\xi)}
where $\mu^2 = m^2 + {\bf p}^2$ and $\nu = \beta \omega/ 2\pi$.
 Starting from \onx\ 
we have the explicit representation as a continuous frequency integral 
\eqn\expl{
{\widetilde \CG}^E (\Delta \tau) =
 \int {d{\bf p} \over (2\pi)^{d-2}} e^{i{\bf p}\cdot\Delta
{\bf y}} \int_0^\infty {d\nu \over \pi^2} \;\cosh[ \nu(\Delta \theta - \pi 
)] \;K_{i\nu} (\mu\,\xi) \;K_{i\nu} (\mu\,\xi')\;.}
The integrals can now
 be evaluated (see for example Appendix A of \refs\rroberto)
to obtain
\eqn\lal{
\CG^E (\Delta \tau) = {1\over 2\pi} \left({m \over 2\pi \Delta s}\right)^{d-2
\over 2} \;K_{d-2 \over 2} (m\Delta s)\;,}
where $\Delta s$ is the geodesic distance on ${\bf R}^d$. The final
result is the Green's function of the Laplacian on ${\bf R}^d$, i.e.
the Hartle--Hawking Euclidean Green's function. Thus, we have shown by
explicit calculation that a real-time thermal correlator with continuous
spectrum at the horizon is the analytic continuation of an Euclidean
correlator that is  perfectly
smooth in the vicinity of $r=r_0$.   We can now
continue  \lal\ back
 to Lorentzian signature and check the large time asymptotics
of the correlator. Consider the geodesic distance between two points 
in \rin\ with equal values of $\xi$ and ${\bf y}$ and $\Delta \tau < \beta$.
 We have 
$\Delta s = 2\xi \sin (\pi \Delta \tau /\beta)$. Performing the analytic
continuation $\Delta s  \rightarrow -2i\,\xi \sinh(\pi\Delta t/
\beta)$. Hence, at large $\Delta t$ the real-time correlator vanishes 
exponentially and   shows no Poincar\'e recurrences. 

Because of the exponential barrier at large $r_*$,
 the dissipation of perturbations located at $r\approx r_0$ 
occurs mostly towards the horizon, 
 with a   lifetime  given by   
\eqn\ltmass{
|{\rm Im} \,\omega |^{-1}  = {2\beta\over \pi (d-1)} = {8 \over (d-1) g'(r_0)}\;,
}
in the massive case, and 
\eqn\ltmassl{
|{\rm Im} \,\omega |^{-1} =
 {2\beta \over \pi (d-2)} = {8 \over (d-2) g'(r_0)}\;.
}
in the massless case. 
We see that the characteristic time scales are independent of the non-zero mass and
are controlled by the Hawking temperature of the black hole. 
  It should be noted however that \ltmass, \ltmassl\ must be viewed only
 as order-of-magnitude
estimates of the true black hole's  damping frequencies.
 Since the Rindler approximation regards the horizon as flat ${\bf R}^d$, results  
are not accurate on scales of the Schwarzschild radius, which can be
of the same order of magnitude as $|{\rm Im} \,\omega\,|^{-1}$.  

\subsec{The Brick Wall}

\noindent

We have traced the absence of recurrences to the peculiar topological properties of the
Euclidean manifold $X$. This special topology guarantees the emergence of a
classical contribution to the entropy \refs\rGH. At  the same time,
it renders the analytic continuation to real time completely blind to the
spacing of black hole energy levels. Hence, 
the restoration of the recurrences depends on dynamical effects that change
the topology of $X$. This conclusion holds for any static horizon, including
de Sitter in the static patch, and implies that the details of unitary time
evolution can only be retrieved by new nonperturbative effects.  

The recurrences can be restored if we give up the smooth nature of $X$ and change
its topology  in an
{\it ad hoc} way. For example,
we may cut out the manifold at a distance $\varepsilon$ from the horizon, imposing
a Dirichlet boundary condition on fields, i.e. 't Hooft's brick-wall model \refs\rbrick.
This reflecting boundary condition  makes the domain of $r_*$ compact, and the
 Schr\"odinger problem \veun\
yields  a discrete  frequency spectrum.

The Euclidean cutoff manifold $X_\varepsilon$ has now cylindrical
topology  and  no classical entropy of
Hawking--Gibbons type. The leading contribution to
 the entropy arises at one-loop level and diverges as $S(X_\varepsilon) \sim
A_H /\varepsilon^{d-2}$. If this is equated with the Bekenstein--Hawking entropy,
we have an {\it a posteriori} adjustment of
the phenomenological parameter, $\varepsilon$.
The effective potential \univh\  for low energy excitations has now discrete
spectrum and the recurrence index ${\overline L} \sim \exp(-S(X_\varepsilon)) $
 has the right order of magnitude.
Of course, such modifications of the boundary conditions are not consistent with the
systematics of the semiclassical expansion. Rather, one should find well-defined
nonperturbative corrections that  impose such effective boundary conditions. From
our discussion we know that they should involve topology-changing processes, and
some candidate configurations will be described in the next section.

\newsec{Poincar\'e Recurrences and Topological Diversity}

\noindent

The recurrence index was estimated as $\bL \sim \exp (-S)$  for systems
with discrete spectrum. This estimate yields $\bL \sim \exp(-A_H /4G_{\rm N})$
for a system whose entropy is dominated by a black hole phase. 
 However, a direct analysis of correlation functions in the background of
a black hole yields $\bL =0$ in perturbation theory. In this respect, the
recurrence index $\bL$ seems to behave quite differently from the partition
function $Z(\beta)$.  

  Nonperturbative effects  of  
$\CO(e^{-1/G_{\rm N}})$  arising in the semiclassical expansion must
effectively discretize the energy spectrum in order to capture the
phenomenon of the recurrences.  It is unlikely that string 
D-instanton corrections in the background \eucl\  
would achieve this goal, at least within the dilute-gas
regime.  
Large-scale fluctuations of the gravitational background that change the
global topology of \eucl\  might be strong enough \refs\rmaldas. 
 Consider, for example, tunneling transitions between \eucl\ and
other   backgrounds with  
discrete spectrum of excitations in perturbation theory. In this case 
we may obtain a nonvanishing
result for the recurrence index $\bL$ from the contribution of these
backgrounds with discrete spectrum.  In this section we obtain a quantitative
estimate of these ``instanton effects"
and we discuss their physical interpretation. 

Let us consider a set of perturbative Euclidean backgrounds $X_\alpha$ that
share the same thermal boundary conditions. The characteristic example of
this in the black-hole applications
 is the substitution of the black-hole metric
by the vacuum metric with a gas of
 perturbative thermal particles at temperature
$1/ \beta$. In the AdS/CFT context, they correspond to different large-$N$
master fields of the CFT that contribute at the same temperature.  
In this situation the total measure $d\mu(\phi, \dots)$ splits into 
the different 
perturbative measures $d\mu_\alpha$ around each background,
\eqn\split{
\int d\mu \rightarrow \sum_\alpha e^{-I_{cl} (X_\alpha)} \int d\mu_\alpha\;,}
where we have explicitly included the  suppression factor by
the {\it classical} Euclidean action of each background.  
The total 
Euclidean correlator is then
 given by an average over the different backgrounds. Applying the substitution \split\
to \siete\ we find   
\eqn\cinco{
G^E(\tau, \tau') = { \sum_\alpha e^{-I(X_\alpha)} \;
\CG^E_\alpha (\tau, \tau') \over \sum_\alpha e^{-
I(X_\alpha)}}.}
Here $I(X_\alpha) = I_{cl} \,(X_\alpha) + \CO(\lambda^0)$
 denotes the perturbative expansion of the effective action
around the background manifold  $X_\alpha$.  
 
Real-time
response functions,  denoted collectively by $G(t)$,  are
 defined through the analytic continuation in the time arguments $\tau, \tau'$  
 as before:  
\eqn\ocho{
G(t) = { \sum_\alpha e^{-I(X_\alpha)} \;\CG_\alpha (t) \over \sum_\alpha e^{-
I(X_\alpha)}}\;.}
This expression applies to thermal correlators  either of type  \genco\ or
of type \lrt, provided the gravitational counterparts $\CG_\alpha (t)$ are 
defined accordingly.  
From this expression we can calculate the ``instanton" approximation to 
$\bL$ with the result
\eqn\ianp{
\bL_{\rm inst} = {\sum_{\alpha,\beta} e^{-I(X_\alpha) - I(X_\beta)} \;
\CG_\alpha (0) \,\CG_\beta (0)^* \;\;{\overline \CL}_{\alpha\beta} \over
\sum_{\alpha,\beta} e^{-I(X_\alpha) - I(X_\beta)} \; 
\CG_\alpha (0) \,\CG_\beta (0)^*}\;,}
where we have defined the time averages of the gravity Green's functions: 
\eqn\mixx{
{\overline \CL}_{\alpha\beta} \equiv \lim_{T \to \infty} \;{1\over T} \,\int_0^T dt \,
\CL_{\alpha\beta} \,(t) \;, \qquad \CL_{\alpha\beta} \,(t) \equiv  
{\CG_\alpha (t) \;\CG_\beta (t)^* \over \CG_\alpha (0) \;\CG_\beta (0)^*}\;.}

The zero-time Green's functions $\CG_\alpha (0)$ are of $\CO(1)$ in the semiclassical
limit. Hence, the denominator of \ianp\ is dominated by the manifold
 with largest 
 partition function at each temperature. We shall denote
this thermodynamically-dominating manifold   $X$.  
On the other hand, the numerator is 
determined by the competition between the thermodynamical
partition function and the time averages ${\overline \CL}_{\alpha \beta}$. At any rate,
the time averages vanish identicaly for any black-hole background. Hence, the background
that dominates the numerator of \ianp, denoted $Y$, does not contain horizons. 
In this way we 
arrive at the final estimate:
\eqn\fines{
\bL_{\rm inst} \approx   
  e^{-2\,\Delta I} \;\left|C_{Y/X}\right|^2 
\,\cdot \, {\overline \CL}_Y   
\;,}  
where we denote  
 $$
\Delta I = I(Y) - I(X) \;\;\;\;\;\;{\rm and}\;\;\;\;\; C_{Y/X} = {\CG_Y (0) \over 
\CG_X (0)}\;.
$$  

The value of ${\overline \CL}_Y$ is controlled by the eigenvalue spacing of
$\CH(Y)$, the perturbative Hamiltonian of gravitational fluctuations in the
background $Y$. This background has no horizons 
and $\CH(Y)$ has discrete spectrum.
Applying the general discussion of Section 2 we obtain ${\overline \CL}_Y \sim
\exp(-S(Y))$. Hence, the instanton approximation to $\bL$ agrees exactly with
the bound \fin\ when $X=Y$, i.e.
 when the thermodynamical free energy is dominated
by a thermal gas with no black holes.

 At sufficiently large temperatures black holes
will be dominating and $X\neq Y$. In this case,  
 the leading exponential suppression of $\bL_{\rm inst}$
is of order 
 $\exp (-2\Delta I) \sim \exp(-1/G_{\rm N})$, 
and the result of \fin\ is recovered
in order of magnitude. 
The agreement is exact in the strict $G_{\rm N} \rightarrow 0$ limit (or $N\rightarrow
\infty$ in the dual CFT), where the entropy is no more finite and the bound itself vanishes.

 We may ask  whether the agreement between \fin\ and \fines\ is
exact also  at the level of  $\CO(\exp(-1/G_{\rm N}))$ accuracy  
 in the exponential suppression. 
 In order to
parametrize the  success of the instanton approximation to $\bL$
 we define the relative error of the exponent 
\eqn\relat{
\eta \equiv {S(X) - 2\Delta I \over S(X)}\;.}
The instanton approximation would be succesful if this quantity turned out to
be zero in the leading $\CO(1/G_{\rm N})$ approximation. 
We now check the value of $\eta$ with an explicit computation in two
characteristic
examples. 

\subsec{Schwarzschild Black Holes in a Box}

\noindent

The simplest system of a black hole in finite volume is that of a
Schwarzschild solution with reflecting boundary conditions on a sphere
of radius $R$. It has the problem that the mathematical characterization
of the box is very unphysical. In addition, it lacks a known holographic
dual. However, we can assume that such a quantum mechanical description
exists and then we can calculate $\eta$ in the semiclassical approximation,
following York, \refs\ryork.  

Let us consider the canonical equilibrium of Schwarzschild black holes with
Euclidean metric
\eqn\mets{
ds^2 = \left(1-(r_0 /r)^{d-3} \right) \,d\tau^2 + {dr^2 \over 
\left(1-(r_0 /r)^{d-3} \right)} + r^2 \; d\Omega_{d-2}^2
}
in spacetime dimensions $d\geq 4$ with a spherical boundary at $r=R$
kept at fixed temperature $1/\beta$. The period of $\tau$ is fixed by
the requirement of smoothness at $r=r_0$ and is given by $\tau \equiv
\tau + \beta_\tau$, where
\eqn\betat{
\beta_\tau = {4\pi \over d-3} \,r_0\;.}
The inverse local temperature at the boundary $\beta$ is the 
proper size of the thermal circle at $r=R$, i.e.
\eqn\propp{
\beta = \beta_\tau \;\sqrt{1-(r_0 /R)^{d-3}} ={4\pi r_0 \over d-3}
\,\sqrt{1-(r_0 /R)^{d-3}}\;.}
This expression shows that such black holes have positive specific heat,
$d\beta/dr_0 <0$,  
for 
$$
r_0 > R\,\left({2\over d-1}\right)^{1\over d-3}
\,,$$
 and that there is
a minimum temperature for equilibrium in the box. 
Using now the standard expressions for the ADM mass and Bekenstein--Hawking
entropy:
\eqn\usual{
M= {(d-2) \,{\rm vol}\,({\bf S}^{d-2}) \over 16\pi G_{\rm N}} \;r_0^{d-3}\;,
\qquad S= {{\rm vol}\,({\bf S}^{d-2}) \over 4 G_{\rm N}  }\,r_0^{d-2}\;,
}
we can obtain the free energy $I= \beta \,M -S$, or
\eqn\act{
I_{\rm bh}= {{\rm vol}\,({\bf S}^{d-2}) \over 4 G_{\rm N}} \;r_0^{d-2}\;\left[
{d-2 \over d-3} \,\sqrt{ 1-(r_0 /R)^{d-3}} -1 \right]\;.}
With this normalization, the free energy of a pure radiation ball is
$I=0$ to order $\CO(1/G_{\rm N})$. The black hole in equilibrium with
radiation dominates over the pure radiation state for Schwarzschild
radii above the critical nucleation radius 
\eqn\domm{
r_0 > r_{\rm nucl} = 
R \,\left({2d-5 \over (d-2)^2}\right)^{1\over d-3} 
\;.}
In this example, the background $X$ is given by a black hole  
for temperatures above the nucleation temperature. The background $Y$
continues to be the flat-space box. 
Here $\Delta I = -I_{\rm bh}$ and the mismatch parameter,    
\eqn\recuin{
\eta =-1 + {2d-4 \over d-3} \,\sqrt{1-(r_0 /R)^{d-3}} 
\,,} 
varies between $\eta=1$ at the nucleation temperature, down to
$\eta = -1$ at infinite temperature. We see that the instanton evaluation
of $\bL$ is generically different from the direct quantum mechanical
estimate. At the particular Schwarschild radius
\eqn\part{
r_0  = R \left({(3d-7)(d-1) \over 4 (d-2)^2}\right)^{1\over d-3} 
}
we have $\eta =0$.

\subsec{AdS Black Holes}

\noindent

A more rigorous example in which the boundary is fully specified by
a gravitational interaction is the case of black holes in AdS space.
This is more interesting, because we may regard the correlators $G(t)$
as defined non-perturbatively in the dual CFT theory on the boundary. 
In this case, for local operators in the bulk of AdS, the corresponding
operators in the CFT are non-local. This is not a problem, since the
general considerations of Section 2 make no assumptions about the
 spatial locality of the operators $A$ and $B$. 

Recall the computation of the  free-energy difference  between  the large
${\rm AdS}_{d}$ black hole, which we denote $X$,  
 and the AdS vacuum, denoted $Y$,  in the classical gravity
approximation (c.f. \refs\rwithp),  
\eqn\est{
\Delta I = I(Y) - I(X) = {{\rm vol}({\bf S}^{d-2}) \over 4 \,G_{\rm N}} \,
{r_0^{d} - R^2 r_0^{d-2} \over (d-1)\,r_0^2 + (d-3)R^2}
\;,}
where $r_0$ stands for the horizon radius, related to the inverse temperature
by the formula
\eqn\htt{
\beta= {4\pi R^2 r_0 \over (d-1)\,r_0^2 + (d-3) R^2}
\;.}
The 
 mismatch 
of the instanton method is now given by 
\eqn\rmi{
\eta  = {(d-3) r_0^2 + (d-1)\,R^2 \over (d-1)\,r_0^2 + (d-3) R^2}
\;.}
We find that the disagreement is in general of $\CO(1)$.
 The mismatch between both
estimates is maximal at the Hawking--Page phase transition $r_0 = R$ where the
ratio
$\eta = 1
$. 
At very large temperatures, $\beta\ll R$,  it decreases according to
the law  
\eqn\decc{
\eta = {d-3 \over d-1} + {1\over 2\pi^2 (d-1)} \left({\beta \over R}
\right)^2 + \CO(\beta^4 /R^4)  
\;.}
For  $d=3$
this asymptotic formula is exact, without any $\CO(\beta^4 /R^4)$
 corrections.  We see
that the two estimates of $\bL$  are exponentially different
at any finite temperature above the phase transition. They only agree for $d=3$
in the strictly infinite  
 temperature limit.   

In $d=3$ there is a rich structure of black holes related by $SL(2, {\bf Z})$
transformations \refs\rbtz.
 However, the previous analysis still applies, provided we only
consider non-rotating black holes.  
At zero angular momentum, 
all the extra black holes are subleading,
either to the vacuum AdS at low temperatures,
or to the high-temperature black hole beyond the phase transition. 

The $d=3$ case presents however an interesting peculiarity. Unlike the $d>3$
case, BTZ black holes  do not approach the vacuum in the zero-mass limit. 
 The smallest regular black hole  
 has  zero  mass
in units in which
the energy of the vacuum is the Casimir energy of the Neveu--Schwarz sector, namely     
\eqn\nsch{
E_{\rm NS \,vac} = -{C \over 12 R} 
}
with $C$ the CFT's central charge. The zero-mass black hole in the NS sector
is degenerate with the Ramond vacuum and has zero temperature and metric
\eqn\rvac{
ds^2 = {r^2 \over R^2} \,(-dt^2 + R^2 \,d\varphi^2 ) + R^2 {dr^2 \over r^2}
\;.}
Thus, this metric describes the Ramond vacuum if we assign periodic boundary
conditions to the fermions in the asymptotic spatial cycle. But this
metric  is just the Poincar\'e patch of the  vacuum AdS manifold, so $r=0$ is
a horizon, with the associated problem of the continuous spectrum of modes 
\refs\rstro.  
In addition, this horizon is actually singular on account of the compact nature
of the angle $\varphi$. So, the Ramond sector of the two-dimensional CFT is
an example  where the vacuum manifold has already
the disease of the continuous spectrum\foot{In this case 
one can hope for improvements
of the semiclassical approximation along the lines of \refs\rmathur.}, 
 and 
it seems impossible that large-distance semiclassical effects could restore
the Poincar\'e recurrences.

A similar phenomenon  takes place in AdS spaces of arbitrary dimension at
sufficiently large temperatures. At very small $\beta$, the vacuum ${\rm AdS}_d$ 
manifold with a thermal gas becomes unstable, either by the Jeans instability or
by other stringy effects. In such a situation, we would lack an appropriate $Y$
background to recover a nonvanishing $\bL$.  For example, the   
  standard ${\rm AdS}_5 \times {\bf S}^5$
 model develops thermal tachyons when  the temperature
reaches either the Hagedorn temperature in the bulk $ R/\beta
 \sim (g_s N)^{1/4}$,
or the Jeans temperature $ R/\beta \sim N^{1/5}$ (c.f. \refs{\rusex, \rthresholds,
\rcensor}).\foot{This is just as well, because otherwise the background $Y$ would
dominate the thermodynamics at temperatures $R/\beta \gg N^{1/3}$, having a
non-holographic scaling in the temperature $I_{Y} \sim -(R/\beta)^9$, to be compared
with $I_{X} \sim -N^2 \,(R/\beta)^3$.}  

\subsec{Long and Very Long Time Scales}

\noindent

We have seen that, while the instanton approximation \fines\ to the unitarity 
bound \fin\ is right in order of magnitude, it is unlikely to be exact to 
order $\exp(-1/\lambda^2)$. We can argue this by  noticing that
  $\bL_{\rm inst}
\sim \exp(-2\Delta I)$ is universal for all types of two-point functions
that can be obtained as analytic continuation of the Euclidean Hartle--Hawking
Green's functions. In
particular, it applies to the $\bL_{\rm LR}$ average in the thermo-field
double. However,  
 we know that the direct bounds of Section 2 have a slight dependence on the
class of correlation function considered (compare   
 \fin\ with \ett). 
Furthermore, the evaluation of $\bL_{\rm inst}$ relies on the existence of
an appropriate $Y$ background with discrete spectrum,   
 a requirement that is not met  in the RR sector of BTZ black holes, or even in
general AdS spaces for sufficiently large temperatures.   

\fig{\sl Schematic representation of
the  very long time behaviour of $L(t)_{\rm inst}$ (dark line)
compared to the expected pattern for the exact quantity $L(t)$. The
resurgences of $L(t)_{\rm inst}$ occur with periods of
order $t_H (Y) = \CO(\lambda^0)$ and have amplitude of order $e^{-1/\lambda^2}
\ll 1$. The expectations for the exact CFT, in the
dashed line, are  $\CO(1)$ resurgences with a much larger period
$t_H  \sim e^{1/\lambda^2} \gg t_H (Y)$, corresponding to tiny energy spacings of
order $e^{-1/\lambda^2}$.
 Despite the gross difference
of both profiles, the infinite time average is $\CO(e^{-1/\lambda^2})$
for both of them.
}{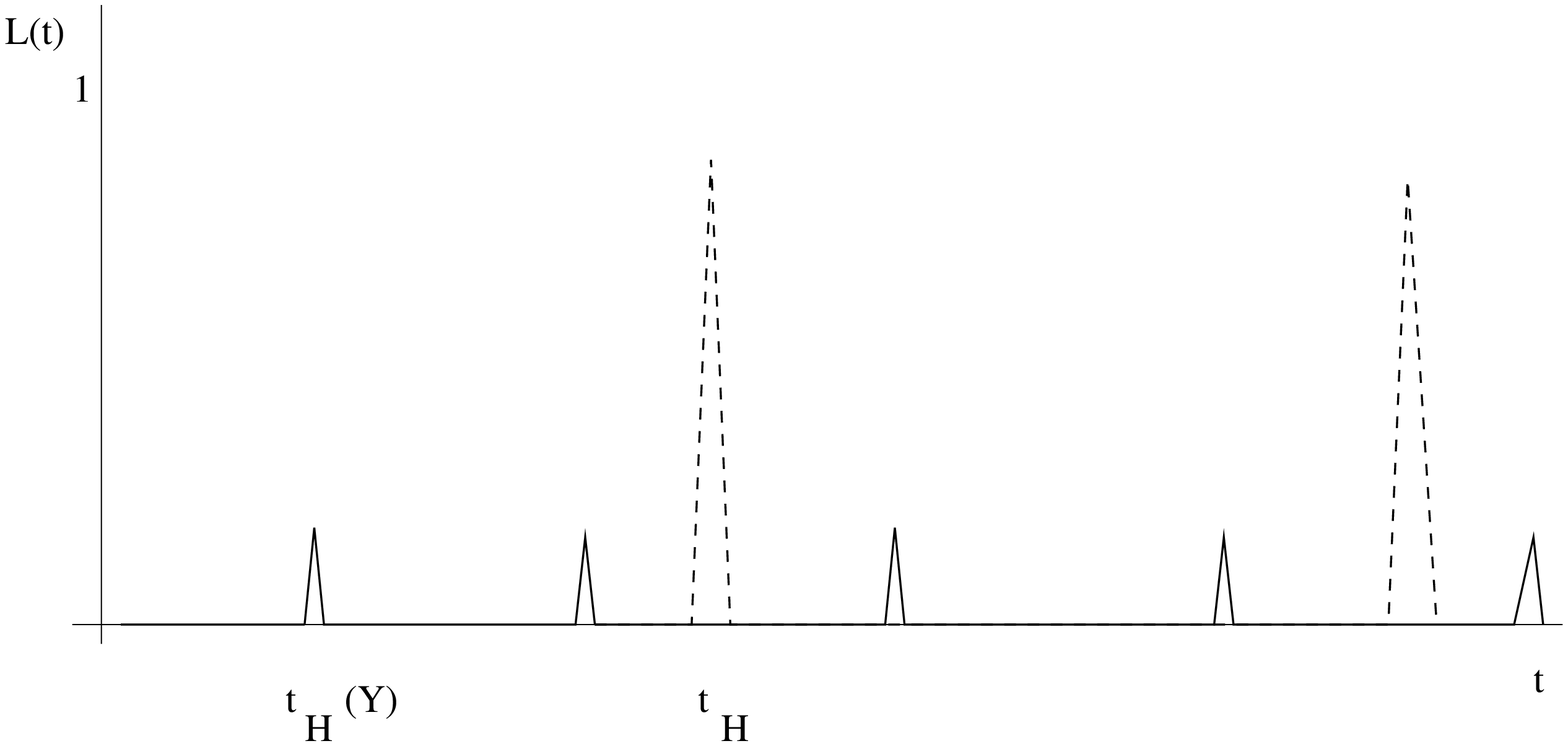}{5truein}

In fact, despite the agreement between \fin\ and \fines,  the instanton processes 
considered so far do not directly address the question of Poincar\'e recurrences.  
Let us 
consider the
instanton prediction for the very long time behaviour of the normalized
ratio $L(t)$, rather than the time average. Since $\lim_{t\to \infty} \CG_X (t)
=0$, we have
\eqn\onces{
L(t)_{\rm inst} \approx e^{-2\Delta I} \;   
\left|C_{Y/X} \right|^2 \;\cdot \CL_Y (t)\;,}
for sufficiently large times. 

In \onces\ the period of quasiperiodicity of $L(t)_{\rm inst}$ is
that of $\CL_Y (t)$. This is controlled by the average eigenvalue
spacing of the perturbative Hamiltonian $\CH(Y)$,  which is of
$\CO(1)$ in the semiclassical expansion in powers of $\lambda$.
  Therefore, the Heisenberg time of $L(t)_{\rm inst}$ scales as 
$t_{H} (Y) =\CO(1)$. On the other hand, the exact CFT has a dense band of
states with spacing of order $\exp(-S) \sim \exp(-1/\lambda^2)$, leading to
a Heisenberg time
    of order $t_H \sim e^{1/\lambda^2}$. Since $t_H (Y) \ll t_H$, the single instanton
approximation does not reflect the correct time scale of the problem.  In addition,
while $\CL_Y (t)$ does show $\CO(1)$ resurgences   on periods
of order $t_{H} (Y)$, the overall prefactor in \onces\ means that
the resurgences of $L(t)_{\rm inst}$  are only of size $e^{-2\Delta I}
\sim e^{-1/\lambda^2} \ll 1$. Thus we conclude that the single instanton approximation
to $L(t)$ has the wrong pattern of Poincar\'e recurrences, both at the
level of the period and also the amplitude.
Interestingly, these two errors tend to cancel one another when
considering the infinite time average $\bL_{\rm inst}$, which simply means
that time averages can be misleading in this particular problem. 

 However, even though the small noise  \onces\ is not directly related to 
the true Poincar\'e recurrences, it  does represent an interesting  correction
to the semiclassical correlators in the black hole background, showing the first
signs of information retrieval. In particular,
its effects are significant much before the Poincar\'e time.  To see this
we can  start from \ocho, before taking the time average, and estimate
$L(t)$ as
\eqn\instt{
L(t)_{\rm inst} = \CL_X \,(t) + e^{-2\Delta I} \;\left|C_{Y/X} \right|^2\;\CL_Y \,(t) +
2\; e^{-\Delta I} \;{\rm Re} \;\left[\;C_{Y/X} \;\CL_{XY}\;(t)\;\right] + \dots}
where the amplitude of $\CL_{XY} \,(t)$ is of order
$
| \CL_X\,(t) \,\cdot \, \CL_Y \,(t)
|^{1/2} \;  
$. 
Approximating the long time behaviour of the black hole correlation functions as 
\eqn\lont{
\CL_X \,(t) \sim \exp (-2\Gamma\,t)\;, }
we find that the $\CO(1)$ ``bumps" of $\CL_Y\,(t)$ represent a significant correction
to $\CL_X \,(t)$ for critical times $t_c$ such that $\exp(-\Gamma t_c) \sim \exp(-\Delta I)$,
that is
\eqn\ttis{
t_c \sim {\Delta I \over \Gamma}\;,}
a very large time for macroscopic black holes, but still exponentially smaller than the
true Heisenberg time of the system. 
In fact, in the strict semiclassical limit, where $\lambda^2 \sim 1/N^2 $
is smaller than any other dimensionless quantity in the system, one has $t_c \gg t_H \,(Y)$,
and the ``instanton noise" \instt\ looks like  a stabilization of the time correlator $L(t)$,
when considering time scales of order $t_c$. 

We have seen in Section 3 that $\Gamma$ is proportional to $ 1/\beta$ up to
   perturbative corrections,
which gives $t_c \sim \beta \,\Delta I$. 
For very massive fields in ${\rm AdS}_3$ black holes,
 corresponding to ${\rm CFT}_2$ operators of very high dimension
$\Delta \sim m\,R \gg 1$, the WKB approximation to the correlator yields   
$\Gamma \sim m\,R/\beta$ and we recover the ``fluctuation time" 
 $t_c \sim \beta \,\Delta I /mR$ of the first paper in Ref. \refs\rshenk.

\newsec{Discussion}

\noindent

Black holes may violate the Poincar\'e theorem on recurrences in two
different ways. First, there will be no recurrences if
 the evaporation process violates
quantum coherence. However, even if we keep the standard unitary formalism
of quantum mechanics, recurrences are averted by the effectively non-compact
phase space of  the black hole horizon. This is exactly what happens in
the semiclassical approximation
to equilibrium gravitational thermodynamics.  The success of the semiclassical
treatment for thermodynamical quantities 
is largely based on the prescription of Hartle and Hawking,
that incorporates horizons by the topological ``no boundary"  condition.     
Here we have shown that the smoothness of the Euclidean black hole manifolds
at the horizon is ultimately responsible for the continuous spectrum in
Lorentzian correlation functions. 

Thus, the horizon ``no boundary" condition is  {\it thermostatic}
in nature  and fundamentally ``coarse grained",   missing important 
information about the details of the thermal ensemble (see also \refs\rmartinec).  
We have seen that the recurrences can be restored by a  singular  
modification of the smooth manifold $X$, such as  the brick-wall
boundary condition of Ref. \refs\rbrick.  
 A more physical boundary condition at the horizon is
only obtained within a nonperturbative formulation of quantum gravity, such
as the AdS/CFT correspondence. In this respect, the infinite storage capacity
of states at the horizon violates the ``stringy exclusion principle" \refs\rstro.  

One limitation of these ideas is the specific nature of concrete AdS/CFT dual pairs, since 
 the correspondence  provides a quantum interpretation of  large
black holes {\it together} with the asymptotic AdS space. It is much less clear
how to generalize these considerations  to large black holes in general spacetimes.
Even the CFT state that corresponds to small, unstable black holes in AdS
has not been identified. 

One point of view would be
that  quantum  properties of a black hole can be
specified by a sort of ``quantum horizon", independently of the particular
asymptotic vacuum where the black hole sits. One can imagine that horizons 
 support a subspace $\CH_H$ of the total Hilbert space of states, and that 
the structure of $\CH_H$ has some universal features.  This is one of the implicit
assumptions of the
  so-called
``stretched horizon"  phenomenological model \refs{\rstre,\rsuscumple, \rlowe}, as well as 
approaches that involve dynamical limits with change of the vacuum (c.f. for example
\refs{\rsfet,\renglert}).
 While semiclassically
all stretched 
horizons are described as thermodynamical 
``hot surfaces" with internal entropy ('t Hooft's
brick-wall model being just the crudest version of the stretched horizon),
the  AdS/CFT correspondence  gives
no concrete evidence that this universality should extend to the detailed quantum
structure. In fact, the confusing situation with the quantum description of
 de Sitter space would speak to the contrary.

Nevertheless, it is possible that part of the quantum structure of stretched horizons
is accesible within a systematic semiclassical expansion in the bulk theory.  
 Time averages of correlators are natural candidates to test this idea, 
being  of order $\exp(-1/\lambda^2)$, where
$\lambda^2 \sim G_{\rm N}$ is the effective expansion parameter of
the gravitational theory (string, M-theory or CFT dual). This suggests
that the recurrences could be restored by summing appropriate instantons.
We have examined one class of instanton transitions proposed in \refs\rmaldas,
in terms of large-scale topology-changing fluctuations of the geometry. 
Although the expected time-averaged bound is obtained  in order of magnitude, there are
reasons to believe that the instanton approximation fails to capture the
Poincar\'e recurrences.
 Specifically,  both the time scale of correlator ``bumps"  and their 
amplitude are much smaller than expected.  
While this class of instanton corrections gives an interesting long-time noise to
the correlation functions, foretelling the restoration of unitarity,
 the large recurrences still stay out of reach of these  
particular semiclassical processes.  

These considerations show the inherent limitations of the ``master field" approximation
when it comes to reveal fine details of the quantum mechanical system. Incidentally, this 
might    be important
in related contexts, such as the program of extracting
 information about spacelike singularities
from thermal CFT correlators, c.f. \refs{\rshenk, \rberk}.

It would be useful to have
a general picture  of how recurrences are to be retrieved from a given microscopic model
of the
stretched horizon, independently of the particular large-distance asymptotics
of the whole system. Our results indicate that large-scale instanton
contributions are not enough. Perhaps instanton  effects on shorter
scales can be identified as the main actors. In this respect, it is
worth pointing out that the smoothness of the Euclidean black-hole
saddle point is the result of a fine-tunning between the curvature and
the temperature identification. If back-reaction effects induce temperature
fluctuations this translates into conical singularities in $X$. In string
theory this could bring interesting dynamics of closed-string tachyon condensation
(c.f. \refs{\raps,\rdabho, \rcensor, \rhaghor, \rolddabho}).

\vskip0.2cm

\noindent {\bf Acknowledgements}

\vskip 0.1cm

\noindent

We would like to thank O. Aharony, L. Alvarez-Gaum\'e,  T. Banks,
K. Bardakci, M. Berkooz, S. Elitzur, C. G\'omez,  
J. Maldacena,   E. 
Martinec, B. Pioline,  S. Shenker   and S. Solodukhin  
for useful discussions. The work of J.L.F.B. was partially supported by MCyT
 and FEDER under grant
BFM2002-03881 and
 the European RTN network
 HPRN-CT-2002-00325. The work of E.R. is supported in part by the
Miller Foundation, the
BSF-American Israeli Bi-National Science Foundation, The Israel Science
Foundation-Centers of Excellence Program, The German-Israel Bi-National
Science Foundation and the European RTN network HPRN-CT-2000-00122.

\listrefs

\bye